\documentclass[lettersize,journal]{IEEEtran}
\usepackage{amsmath,amsfonts}
\usepackage{algpseudocode}
\usepackage{algorithm}
\usepackage{array}
\usepackage[caption=false,font=normalsize,labelfont=sf,textfont=sf]{subfig}
\usepackage{textcomp}
\usepackage{stfloats}
\usepackage{url}
\usepackage{verbatim}
\usepackage{graphicx}
\usepackage{cite}
\usepackage{bbding}
\usepackage{pifont}
\usepackage{makecell}
\usepackage{multirow}
\newcommand{\cmark}{\ding{51}}%
\newcommand{\xmark}{\ding{55}}%
\begin{document}

\title{Layer-Adapted Implicit Distribution Alignment Networks for Cross-Corpus Speech Emotion Recognition}

\author{Yan Zhao, Yuan Zong$^*$,~\IEEEmembership{Member,~IEEE}, Jincen Wang, Hailun Lian, Cheng Lu, Li Zhao, and Wenming~Zheng$^*$,~\IEEEmembership{Senior~Member,~IEEE}
\thanks{Y. Zhao, H. Lian and L. Zhao are with the Key Laboratory of Child Development and Learning Science of Ministry of Education, and also with the School of Information Science and Engineering, Southeast University, Nanjing 210096, China. (e-mail: \{zhaoyan, 230208573, zhaoli\}@seu.edu.cn).}
\thanks{Y. Zong, J. Wang, C. Lu and W. Zheng are with the Key Laboratory of Child Development and Learning Science of Ministry of Education, Southeast University, Nanjing 210096, China, and also with the the School of Biological Science and Medical Engineering, Southeast University, Nanjing 210096, China. (e-mail: \{xhzongyuan, 220222338, cheng.lu, wenming\_zheng\}@seu.edu.cn).}
\thanks{* indicates the corresponding authors.}
}

\markboth{Journal of \LaTeX\ Class Files,~Vol.~14, No.~8, August~2023}%
{Shell \MakeLowercase{\textit{et al.}}: A Sample Article Using IEEEtran.cls for IEEE Journals}

\maketitle

\begin{abstract}
In this paper, we propose a new unsupervised domain adaptation (DA) method called layer-adapted implicit distribution alignment networks (LIDAN) to address the challenge of cross-corpus speech emotion recognition (SER). LIDAN extends our previous ICASSP work, deep implicit distribution alignment networks (DIDAN), whose key contribution lies in the introduction of a novel regularization term called implicit distribution alignment (IDA). This term allows DIDAN trained on source (training) speech samples to remain applicable to predicting emotion labels for target (testing) speech samples, regardless of corpus variance in cross-corpus SER. To further enhance this method, we extend IDA to layer-adapted IDA (LIDA), resulting in LIDAN. This layer-adpated extention consists of three modified IDA terms that consider emotion labels at different levels of granularity. These terms are strategically arranged within different fully connected layers in LIDAN, aligning with the increasing emotion-discriminative abilities with respect to the layer depth. This arrangement enables LIDAN to more effectively learn emotion-discriminative and corpus-invariant features for SER across various corpora compared to DIDAN. It is also worthy to mention that unlike most existing methods that rely on estimating statistical moments to describe pre-assumed explicit distributions, both IDA and LIDA take a different approach. They utilize an idea of target sample reconstruction to directly bridge the feature distribution gap without making assumptions about their distribution type. As a result, DIDAN and LIDAN can be viewed as implicit cross-corpus SER methods. To evaluate LIDAN, we conducted extensive cross-corpus SER experiments on EmoDB, eNTERFACE, and CASIA corpora. The experimental results demonstrate that LIDAN surpasses recent state-of-the-art explicit unsupervised DA methods in tackling cross-corpus SER tasks.
\end{abstract}

\section{Introduction}

\IEEEPARstart{T}{he} research of speech emotion recognition (SER) aims to equip computers with the ability to automatically understand the emotional states conveyed in speech signals, e.g., \textit{Happiness}, \textit{Sadness}, and \textit{Surprise}~\cite{el2011survey,akccay2020speech}. If the computers possessed this capability, their interactions with humans would become more natural. Consequently, SER has emerged as a prominent and intriguing topic within the fields of affective computing, speech signal processing, and human-computer interaction over the past few decades. One of the earliest works on SER can be traced back to the study of Williams and Stevens~\cite{williams1972emotions}, in which they investigated whether acoustic parameters extracted from speech signals recorded under different emotional states exhibited distinct characteristics. Since then, significant efforts have been devoted to SER, leading to numerous effective SER methods~\cite{schuller2003hidden,wu2011automatic,huang2014speech,zhang2019spontaneous,issa2020speech,lu2022domain,lu2022speech}.

It is worth noting that most of these SER methods often assume that the feature distributions of the training and testing speech signals follow the same or similar distribution, which allows them to achieve promising performance. However, in practical application scenarios, this assumption may be easily violated. For example, the training and testing speech signals encountered by the SER system may be most likely recorded by quite different acoustic sensors or in different languages. This introduces a new task in SER known as cross-corpus SER~\cite{schuller2010cross}. Formally speaking, the source (training) and target (testing) speech signals in cross-corpus SER tasks belong to different corpora, which is quite different from the conventional SER setting. Moreover, only the emotion labels of the training speech signals are given, while the testing ones are completely unavailable. Consequently, cross-corpus SER is obviously a more challenging task than the conventional one, and therefore the performance of these originally well-performing SER methods may decrease sharply, which does not satisfy the requirements of practical applications. 

Researchers have been aware of this issue in SER and have tried to address it from different angles. For example, Schuller et al.~\cite{schuller2010cross} proposed three different feature normalization strategies, including speaker normalization (SN), corpus normalization (CN), and speaker-corpus normalization (SCN), to overcome the issue of corpus variance in cross-corpus SER tasks. On the other hand, some researchers found that cross-corpus SER tasks can be seen as a typical unsupervised domain adaptation (DA) problem~\cite{wang2018deep,zhuang2020comprehensive}. Therefore, ideas widely used in the research of unsupervised DA have been employed to develop subspace or deep learning-based DA methods to alleviate the feature distribution mismatch between the source and target speech signals. For example, Hassan et al.~\cite{hassan2013acoustic} proposed an importance-weighted support vector machine (IW-SVM) to address the cross-corpus SER problem. They applied the importance weighted transfer learning method, e.g., kernel mean matching (KMM)~\cite{gretton2009covariate}, to narrow down the maximum mean discrepancy (MMD)~\cite{borgwardt2006integrating} distance between the weighted source and target speech feature sets. In the work of~\cite{zhang2021cross}, Zhang et al. presented a novel unsupervised DA framework, called joint distribution adaptive regression (JDAR), for cross-corpus SER. JDAR aims to find a corpus-invariant common subspace to bridge samples from different speech emotion corpora by minmizing the first-order statistical moment, i.e., mean value, of the marginal and conditional feature distributions of source and target speech samples. Recently, Zhao et al.~\cite{zhao2022deep} proposed a novel deep unsupervised DA method called deep transductive transfer regression networks (DTTRN), which includes a major module based on multi-kernel MMD for adapting source and target speech feature distributions. 

From the aforementioned works, it is evident that these methods can be categorized as \textit{Explicit Methods} for addressing the issue of cross-corpus SER. This categorization is attributed to the assumption underlying their design, which considers the distributions of the source and target speech feature sets as explicitly pre-defined, e.g., following a Gaussian distribution. Consequently, appropriate statistical moments such as the mean value can be used to characterize the feature distribution in finite or kernel-induced infinite subspaces. The aim is then to minimize the distance between these distributions and thereby mitigate the mismatch between the distributions of source and target speech corpora. However, it is important to note that this assumption of speech samples being distributed following the assumed feature distributions may not hold true in reality. As a result, the chosen statistical moments can not accurately capture the true feature distribution. For this reason, these explicit methods, which aims to minimize the distance between statistical moments, such as MMD, may fail to effectively bridge the gap in feature distribution between the source and target speech corpora.

To overcome the limitations of the explicit methods, researchers have attempted to tackle the issue of cross-corpus SER from a different perspective, which can be summarized as \textit{Implicit Methods}. Their aim is to bridge the gap between the feature distributions of source and target speech signals without relying on specific statistical moments to describe the pre-assumed distribution. One noteworthy example is the work of Abdelwahab and Busso on adversarial multitask learning for cross-corpus SER~\cite{abdelwahab2018domain}. In this work, they introduce an adversarial multitask learning framework to acquire corpus-invariant common representations of emotional speech signals across different speech corpora, rather than first assuming the speech feature distributions and then calculating statistical moment distances. More specifically, to prevent the model from being aware of corpus variance in cross-corpus SER, an additional module called the domain (corpus) classifier is incorporated to alternatively train with the emotion classifier. Subsequent to the success of Abdelwahab and Busso's pioneering work, several variants of adversarial learning models have emerged in recent years for cross-corpus SER tasks. Representative examples of such approaches include adversarial discriminative domain generalization (ADDoG)~\cite{gideon2021improving}, self-supervised adversarial dual discriminator (sADDi)~\cite{latif2022self}, and adversarial domain generalized Transformer (ADoGT)~\cite{gao2023adversarial}. These models have showcased the potential of the adversarial learning framework in capturing the underlying latent representations shared across different speech corpora, surpassing the performance of explicit methods.

Inspired by the success of the adversarial learning-based methods, our previsous work also focused on the research of implicit cross-corpus SER methods and proposed a novel implicit deep unsupervised DA model called deep implicit distribution alignment networks (DIDAN), which was published in ICASSP~2023~\cite{zhao2023deep}. The main contribution of the proposed DIDAN was the design of an implicit distribution alignment (IDA) module guided by the idea of target sample reconstruction, distinguishing it from adversarial learning methods. This module aimed to bridge the feature distribution gap between source and target speech samples without assuming specific feature distributions. In this paper, we extend our previous work, DIDAN~\cite{zhao2023deep}, to a more effective implicit model called layer-adapted implicit distribution alignment networks (LIDAN) for cross-corpus SER. Compared with DIDAN, we introduce a multi-layer version of IDA called layer-adapted IDA (LIDA) for LIDAN, which is comprised of three different modified IDA terms. These terms are deliberately organized across various fully connected layers in LIDAN, in accordance with the enhanced emotion-discriminative capabilities as the layer depth increases. With this enhancement, LIDAN achieves a more effective alleviation of feature distribution mismatch between different speech corpora while simultaneously learning emotion-discriminative and corpus-invariant features for cross-corpus SER. In summary, compared with our conference work, the additional contributions of this paper are three-fold:
\begin{enumerate}
\item We propose LIDAN, an extended version of our previous conference work on DIDAN, which provides a more effective solution to the problem of cross-corpus SER. To the best of our knowledge, DIDAN and LIDAN are the first works that introduce the concept of implicit methods in cross-corpus SER research, avoiding the distribution pre-assumptions required by explicit methods.
\item To enable LIDAN to learn both emotion-discriminative and corpus-invariant features for cross-corpus SER, we have designed a multi-layer version of IDA, LIDA module. This module effectively balances the emotion-discriminative capacity and corpus-invariant ability of different layers in LIDAN, ensuring the alleviation of feature distribution mismatch between diverse speech corpora.
\item We have conducted more extensive cross-corpus SER experiments on three publicly available speech emotion corpora: EmoDB, eNTERFACE, and CASIA. The experimental results demonstrate the effectiveness and superior performance of LIDAN in addressing the challenge of cross-corpus SER.
\end{enumerate}

The remainder of this paper is organized as follows: Section~\ref{sec:method} provides a review of our previous work on DIDAN for cross-corpus SER and subsequently presents the extension from DIDAN to LIDAN. In Section~\ref{sec:exp}, we conduct extensive experiments to evaluate the performance of the proposed LIDAN in tackling cross-corpus SER tasks. Finally, our work is concluded in Section~\ref{sec:conclusion}.

\section{Proposed Method}
\label{sec:method}

\subsection{Preliminary}

In this section, we will provide a detailed explanation of how to extend DIDAN to LIDAN and demonstrate how LIDAN can be utilized to address the problem of cross-corpus SER. To begin with, suppose we have a source speech emotion corpus consisting of $N_s$ samples, denoted as $\mathcal{D}_{s} = \{(\mathcal{X}_{i}^{s}, \mathbf y_{i}^{s})\}_{i=1}^{N_s}$, where $\mathcal{X}_{i}^{s}$ is the Mel-spectrum pseudo image of the $i^{th}$ speech sample and $\mathbf y_{i}^{s} \in \mathbb{R}^{C\times1}$ is its corresponding one-hot label vector generated based on the emotion ground truth, respectively. Herein, $C$ represents the number of emotions involved in the task of cross-corpus SER. Similarly, the unlabeled speech samples from the target corpus can be denoted as $\mathcal{D}_{t} = \{\mathcal{X}_{i}^{t}\}_{i=1}^{N_t}$, with $\mathcal{X}_{i}^{t}$ representing the Mel-spectrum pseudo image of the $i^{th}$ target speech sample and $N_t$ representing the total number of target speech samples.

\begin{figure*}[t]
\centering
\includegraphics[width=0.8\textwidth]{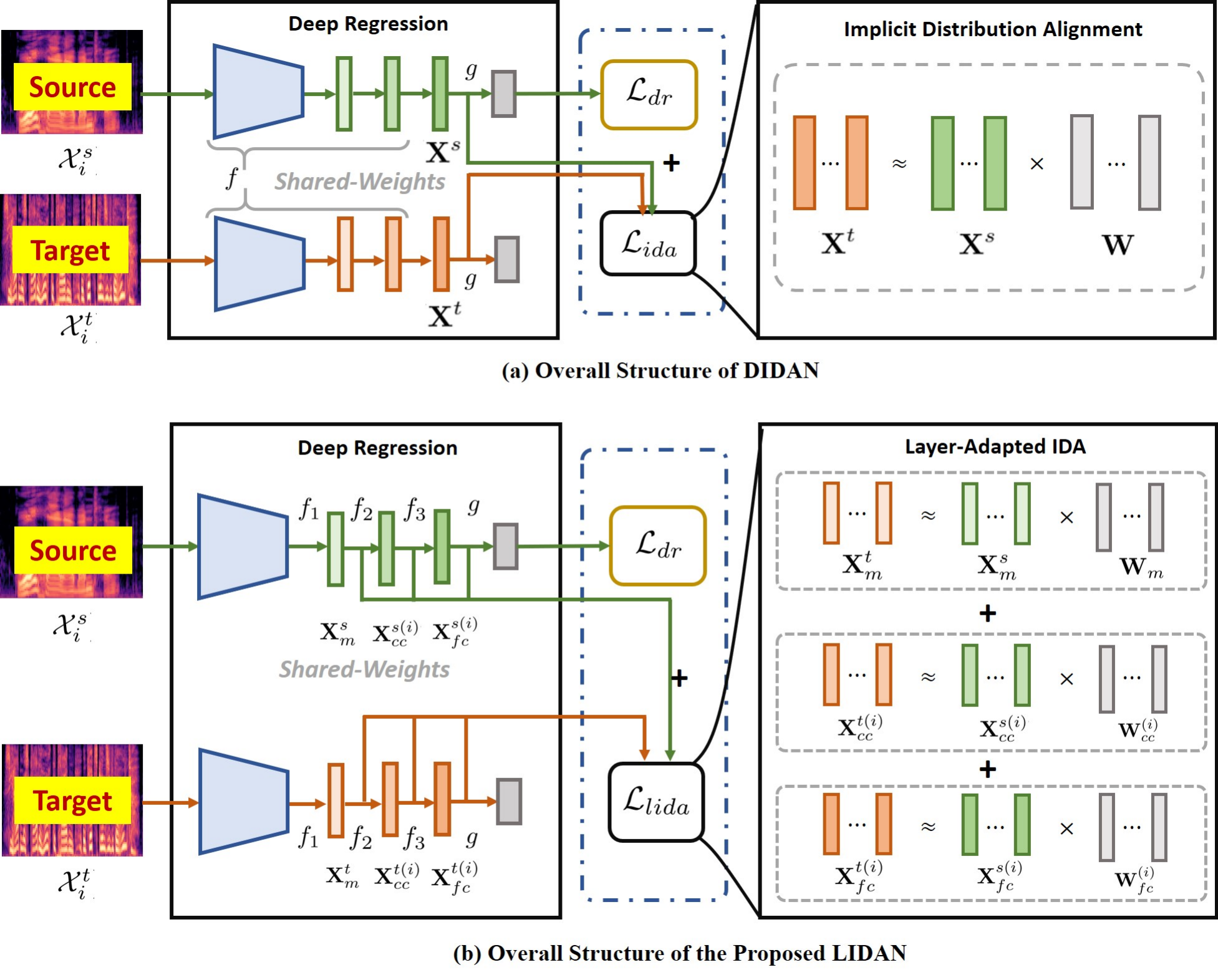}
\caption{Overview Structures of (a) DIDAN~\cite{zhao2023deep} and Its Extended Version, (b) the proposed LIDAN, for Addressing the Issue of Cross-Corpus SER.}
\label{fig:nb1}
\end{figure*}

\subsection{Building DIDAN}

As LIDAN is based on DIDAN, we will first provide a brief overview of DIDAN~\cite{zhao2023deep}. To illustrate its fundamental idea and network structure, we have re-drawn the overall picture of DIDAN in Fig.~\ref{fig:nb1} (a). DIDAN comprises two main modules: the \textit{Deep Regression (DR)} and the \textit{Implicit Distribution Alignment (IDA)}. The DR module is composed of convolutional and fully connected layers. Its purpose is to establish a connection between the source speech Mel-spectrograms and their corresponding emotion labels, enabling DIDAN to possess emotion-discriminative capabilities. In this module, we use $f$ to represent the combination operations of convolution and full connection, which are responsible for feature learning. Additionally, $g$ represents the full connection and softmax operations, which serve as the emotion classifier in DIDAN. The objective function for the DR module is denoted as $\mathcal{L}_{dr}$ and is defined as follows:
\begin{equation}
\mathcal{L}_{dr}=\frac{1}{N_{s}} \sum_{i=1}^{N_{s}} J(g(f(\mathcal{X}_i^s)), \mathbf{y}_{i}^{s}),
\label{eqn:nb1}
\end{equation}
where $J(\cdot)$ denotes the cross-entropy loss function. It is clear to see that by feeding the source speech samples into the DR module and minimizing the aforementioned loss function, the DIDAN can DIDAN can progressively learn to differentiate between various emotional speech signals.

The second module is the IDA, which aims to enable DIDAN to recognize emotions in target speech signals, regardless of the differences between the source and target corpora. Unlike MMD-based methods that focus on measuring and reducing the gap of pre-assumed feature distributions explicitly, the IDA module adopts a target sample reconstruction approach. This approach encourages each target speech feature learned in the last fully connected layer in $f$ to resemble the corresponding source features. Consequently, the DIDAN model is also required to train by minimizing the following loss function:
\begin{equation}
\mathcal{L}_{ida} = \Vert \mathbf{X}^t - \mathbf{X}^s \mathbf{W} \Vert^{2}_{F} + \alpha \Vert \mathbf{W}\Vert_{1}.
\label{eqn:nb2}
\end{equation}

In this equation, $\mathbf{X}^s = [f(\mathcal{X}_1^s), \cdots, f(\mathcal{X}_{N_s}^s)] \in \mathbb{R}^{d\times N_s}$ represents the source speech features, $\mathbf{X}^t = [f(\mathcal{X}_1^t), \cdots, f(\mathcal{X}_{N_t}^t)] \in \mathbb{R}^{d\times N_t}$ represents the target speech features, and $\mathbf{W} = [\mathbf{w}_{1},\cdots,\mathbf{w}_{N_t}] \in \mathbb{R}^{N_s \times N_t}$ is a target sample reconstruction coefficient matrix. Each column of $\mathbf{W}$, denoted as $\mathbf{w}_i$, corresponds to a target speech sample. $\Vert \mathbf{W}\Vert_{1} = \sum_{i=1}^{N_t}\Vert \mathbf{w}_i\Vert_1$ is a $L_1$ norm with respect to the reconstruction coefficient matrix. By minimizing this norm, DIDAN learns a sparse $\mathbf{w}_i$, which means that only a few source samples are involved in reconstructing the $i^{th}$ target sample. Additionally, $\alpha$ is a trade-off parameter to control the spasity of the learned reconstruction coefficient matrix. 

Finally, by combining Eqs.~(\ref{eqn:nb1}) and (\ref{eqn:nb2}), the total loss function for learning DIDAN is obtained. The corresponding optimization problem can be formulated as follows:
\begin{equation}
\min_{\theta_{f}, \theta_{g}, \mathbf{W}} \mathcal{L}_{dr}+\lambda \mathcal{L}_{ida},
\label{eqn:nb3}
\end{equation}
where $\theta_{f}$ and $\theta_{g}$ represent the network parameters corresponding to the operations of $f$ and $g$, respectively, and $\lambda$ is the trade-off parameter that balances the losses corresponding to DR and IDA modules.

\subsection{From DIDAN to LIDAN}

To illustrate the concept of LIDAN and highlight its distinctions from DIDAN, we present the overall structure of LIDAN in Fig.~\ref{fig:nb1} (b). Similar to DIDAN, LIDAN comprises two key components: \textit{Deep~Regression (DR)} and \textit{Layer-Adapted IDA (LIDA)}. These components correspond to two well-crafted loss functions, denoted as $\mathcal{L}_{dr}$ and $\mathcal{L}_{lida}$.

\subsubsection{Deep Regression}

In the case of LIDAN, we rewrite the formulation of $\mathcal{L}_{dr}$ in DR module as follows, which allows us to conveniently describe $\mathcal{L}_{lida}$ subsequently:
\begin{equation}
\mathcal{L}_{dr} = \frac{1}{N_{s}} \sum_{i=1}^{N_{s}} J(g(f_3(f_2(f_1(\mathcal{X}_i^s)))), \mathbf{y}_{i}^{s}),
\label{eqn:nb4}
\end{equation}
where $f_1$, $f_2$, and $f_3$ represent the operations performed by convolutional and full connected layers in LIDAN, respectively. From Fig.~\ref{fig:nb1}, it can be observed that their combination, $f_3(f_2(f_1(\cdot)))$, is equivalent to $f(\cdot)$ described in Eq.~(\ref{eqn:nb1}).

\subsubsection{Layer-Adapted IDA}

The primary enhancement in LIDAN is the modification of the IDA module into the layer-adapted multi-layer version, LIDA. This updated version takes advantage of the hierarchical structure inherent in deep neural networks to address differences in feature distribution across different speech corpora layer by layer. LIDA also incorporates the concept of joint distribution adaptation (JDA)~\cite{long2013transfer} from domain adaptation. Specifically, JDA introduces conditional probability distributions to estimate the distribution gap, resulting in more effective distribution alignment compared to marginal distributions. However, it requires additional steps for accurately predicting labels of originally unlabeled target samples to describe their conditional probability distribution, heavily relying on the discriminative abilities of learned features in the corresponding layers.

Under these considerations, we introduce two conditional modifications of IDA alongside the original IDA to create the LIDA module for LIDAN. These IDA modifications are thoughtfully integrated into different fully connected layers to align with the hierarchical learning nature of deep neural networks, where features become increasingly discriminative from shallow to deep layers. Following this idea, the loss function of LIDA, $\mathcal{L}_{lida}$, is designed as follows:
\begin{eqnarray}
\mathcal{L}_{lida} = \mathcal{L}_{lida_m} + \mathcal{L}_{lida_{(cc)}} + \mathcal{L}_{lida_{(fc)}}.
\label{eqn:nb5}
\end{eqnarray}
As shown in Eq.~(\ref{eqn:nb5}), $\mathcal{L}_{lida}$ is comprised of three distinct modified IDA losses: $\mathcal{L}_{lida_m}$, $\mathcal{L}_{lida_{(cc)}}$, and $\mathcal{L}_{lida_{(fc)}}$. These losses correspond to different fully connected layers within the network, ranging from shallow to deep, as shown in Fig.~\ref{fig:nb1} (b). 

The first component, $\mathcal{L}_{lida_m}$, can be referred to as \textit{Marginal Distribution Guided IDA}. It is formulated as:
\begin{eqnarray}
\mathcal{L}_{lida_m} = \Vert \mathbf{X}_m^t - \mathbf{X}_m^s \mathbf{W}_m \Vert^{2}_{F} + \alpha \Vert \mathbf{W}_m\Vert_{1},
\label{eqn:nb6}
\end{eqnarray}
where $\mathbf{X}_m^s = [f_1(\mathcal{X}_1^s), \cdots, f_1(\mathcal{X}_{N_s}^s)]$ and $\mathbf{X}_m^t = [f_1(\mathcal{X}_1^t), \cdots, f_1(\mathcal{X}_{N_t}^t)]$, and $\mathbf{W}_m \in \mathbb{R}^{N_s\times N_t}$ represents the target sample reconstruction coefficient matrix. It is important to note that $\mathcal{L}_{lida_m}$ is identical to the IDA loss used in DIDAN. This design is made based on the consideration that shallow layers in deep neural networks usually have less discriminative features. Therefore, we have shifted the IDA constraint from the deep layer to the shallow layer of LIDAN, as it does not require any emotion information from unlabeled target speech samples.

The remaining two components, $\mathcal{L}_{lida_{(cc)}}$ and $\mathcal{L}_{lida_{(fc)}}$, are conditional modifications of IDA. They can be referred to as the \textit{Coarse-grained Emotion Label Aware Conditional Distribution Guided IDA} and \textit{Fine-grained Emotion Label Aware Conditional Distribution Guided IDA}, respectively. As illustrated in Fig.~\ref{fig:nb1} (b), these components correspond to the deeper fully connected layers in LIDAN. Their objective is to utilize coarse-grained and fine-grained emotion labels to guide the alignment of conditional feature distributions between the source and target speech samples in their respective layers. 

Specifically, $\mathcal{L}_{lida_{(cc)}}$ corresponds to the first deeper layer. It is worth noting that although the discriminative ability of the features learned in this layer experiences a moderate increase compared to the previous layer, it is not yet sufficiently powerful. Therefore, it is beneficial to alleviate the difficulty of predicting the emotion labels of target samples, which is required for aligning conditional feature distributions. To achieve this goal, each speech sample from both the source and target corpora is assigned a coarse-grained emotion label by combining multiple original emotion labels into a single label. As a result, with the features learned in this layer, predicting the coarse-grained emotion labels of target speech samples becomes notably easier compared to predicting the original fine-grained labels. This approach provides a suitable means of aligning conditional feature distributions in this layer for the cross-corpus SER problem. Under this consideration, we modify the original IDA as follows to serve as 
$\mathcal{L}_{lida_{(cc)}}$:
\begin{eqnarray}
\mathcal{L}_{lida_{(cc)}} = \sum_{i=1}^{C_{cc}}\Vert \mathbf{X}^{t(i)}_{cc} - \mathbf{X}^{s(i)}_{cc}\mathbf{W}_{cc}^{(i)} \Vert^{2}_{F} + \alpha \sum_{i=1}^{C_{cc}} \Vert \mathbf{W}^{(i)}_{cc}\Vert_{1}.
\label{eqn:nb7}
\end{eqnarray}

In Eq.~(\ref{eqn:nb7}), $\mathbf{X}^{s(i)}_{cc} = [f_2(f_1(\mathcal{X}_1^{s(i)})), \cdots, f_2(f_1(\mathcal{X}_{N_{s(i)}}^{s(i)}))]$ and $\mathbf{X}^{t(i)}_{cc} = [f_2(f_1(\mathcal{X}_1^{t(i)})), \cdots, f_2(f_1(\mathcal{X}_{N_{s(i)}}^{t(i)}))]$ represent the source and target speech samples with the $i^{th}$ $(i = \{1, \cdots, C_{cc}\})$ coarse-grained emotion label. Additionally, $\mathbf{W}_{cc}^{(i)} \in \mathbb{R}^{N_{s(i)}\times N_{t(i)}}$ is the corresponding target sample reconstruction coefficient matrix. Here, $N_{s(i)}$ and $N_{t(i)}$, satisfy $N_{s(1)} + \cdots + N_{s(C_{cc})} = N_s$ and $N_{t(1)} + \cdots + N_{t(C_{cc})} = N_t$, representing the number of source and target speech samples belonging to the $i^{th}$ coarse-grained emotion label. Moreover, $C_{cc}$ refers to the number of coarse-grained emotion classes.

$\mathcal{L}_{lida_{(fc)}}$ corresponds to the second deeper fully connected layer, which is positioned close to the prediction layer, in LIDAN. In comparison to the previous layers, this layer undoubtedly learns more distriminative features, enabling LIDAN to effectively align fine-grained emotion label guided conditional feature distributions in this layer. Therefore, we modify $\mathcal{L}_{lida_{(fc)}}$ based on the original IDA as follows:
\begin{eqnarray}
\mathcal{L}_{lida_{(fc)}} = \sum_{i=1}^{C}\Vert \mathbf{X}^{t(i)}_{fc} - \mathbf{X}^{s(i)}_{fc}\mathbf{W}_{fc}^{(i)} \Vert^{2}_{F} + \alpha \sum_{i=1}^{C} \Vert \mathbf{W}_{fc}^{(i)}\Vert_{1},
\label{eqn:nb8}
\end{eqnarray}
where $\mathbf{X}^{s(i)}_{fc}$ and $\mathbf{X}^{t(i)}_{fc}$ denote the learned features of source and target speech samples in this layer. They can be represented as $\mathbf{X}^{s(i)}_{fc} = [f_3(f_2(f_1(\mathcal{X}_1^{s(i)}))), \cdots, f_3(f_2(f_1(\mathcal{X}_{N_{s(i)}}^{s(i)})))]$ and $\mathbf{X}^{t(i)}_{fc} = [f_3(f_2(f_1(\mathcal{X}_1^{t(i)}))), \cdots, f_3(f_2(f_1(\mathcal{X}_{N_{s(i)}}^{t(i)})))]$, respectively. Furthermore, $\mathbf{W}_{fc}^{(i)} \in \mathbb{R}^{N_{s(i)}\times N_{t(i)}}$ represents the target sample reconstruction coefficient matrix associated with the source and target speech samples belonging to the $i^{th}$ $(i = \{1, \cdots, C\})$ fine-grained emotion label (original emotion label). 

\begin{algorithm}[t!]
\renewcommand{\algorithmicrequire}{\textbf{Input:}}
\renewcommand{\algorithmicensure}{\textbf{Output:}}
\caption{Updating Rule for Training LIDAN}
\label{al1}
\begin{algorithmic}[1]
\Require Source speech sample set $\mathcal D_{s} = \{(\mathcal{X}_{i}^{s}, \mathbf y_{i}^{s})\}_{i=1}^{N_s}$, target speech sample set $\mathcal D_{t} = \{\mathcal{X}_{i}^{t} \}_{i=1}^{N_s}$, preset fixed trade-off parameters $\{\lambda, \alpha\}$, and maximal iteration $K_{Iter}$.
\Ensure Network parameters $\{\theta_{f_1}, \theta_{f_2}, \theta_{f_3}, \theta_g\}$, target sample reconstruction matrices $\mathbf{W}_m$, $\{\mathbf{W}_{cc}^{(i)}\}_{i=1}^{C_{cc}}$, $\{\mathbf{W}_{fc}^{(i)}\}_{i=1}^{C}$, and target pseudo emotion labels $\{\mathbf y_{i}^{t}\}_{i=1}^{N_t}$.
\State Initialize target pseudo emotion labels $\{{\mathbf y_{i}^{t}}^{(0)}\}_{i=1}^{N_t}$, target sample reconstruction matrices $\mathbf{W}_m^{(0)}$, $\{{\mathbf{W}_{cc}^{(i)}}^{(0)}\}_{i=1}^{C_{cc}}$, $\{{\mathbf{W}_{fc}^{(i)}}^{(0)}\}_{i=1}^{C_{cc}}$, network parameters $\{\theta_{f_1}^{(0)}, \theta_{f_2}^{(0)}, \theta_{f_3}^{(0)}, \theta_g^{(0)}\}$, and iteration index $k=0$, 
\For {$k$ $\mathbf{in}$ $K_{Iter}$}
\State Calculate the totoal loss function $\mathcal{L}_{total}^{(k)} = \mathcal{L}_{dr}^{(k)} + \lambda\mathcal{L}_{lida}^{(k)}$ according to target pseudo emotion labels $\{{\mathbf y_{i}^{t}}^{(k)}\}_{i=1}^{N_t}$, target sample reconstruction matrices $\mathbf{W}_m^{(k)}$, $\{{\mathbf{W}_{cc}^{(i)}}^{(k)}\}_{i=1}^{C_{cc}}$, $\{{\mathbf{W}_{fc}^{(i)}}^{(k)}\}_{i=1}^{C_{cc}}$, and Eqs.~(\ref{eqn:nb4}), (\ref{eqn:nb6}), (\ref{eqn:nb7}), and (\ref{eqn:nb8}), and update model parameters:

\State \hspace{0.5cm} $\theta_{f_1}^{(k+1)} \leftarrow \theta_{f_1}^{(k)} - \mu \frac{\partial \mathcal{L}_{total}^{(k)} }{\partial \theta_{f_1}} $;
\State \hspace{0.5cm} $\theta_{f_2}^{(k+1)} \leftarrow \theta_{f_2}^{(k)} - \mu \frac{\partial \mathcal{L}_{total}^{(k)} }{\partial \theta_{f_2}} $;
\State \hspace{0.5cm} $\theta_{f_3}^{(k+1)} \leftarrow \theta_{f_3}^{(k)} - \mu \frac{\partial \mathcal{L}_{total}^{(k)} }{\partial \theta_{f_3}} $;
\State \hspace{0.5cm} $\theta_{g}^{(k+1)} \leftarrow \theta_{g}^{(k)} - \mu \frac{\partial \mathcal{L}_{total}^{(k)} }{\partial \theta_{g}} $;
\State Update target sample reconstruction matrices:
\State \hspace{0.5cm}$\mathbf{W}_m^{(k+1)} = \arg\min_{\mathbf{W}_m} \Vert \mathbf{X}_m^t - \mathbf{X}_m^s \mathbf{W}_m \Vert^{2}_{F}$
\Statex \hspace{0.5cm} \qquad \qquad \qquad $ + \alpha \Vert \mathbf{W}_m\Vert_{1}$;
\State \hspace{0.5cm}${\mathbf{W}_{cc}^{(i)}}^{(k+1)} = \arg\min_{\mathbf{W}_{cc}^{(i)}} \Vert \mathbf{X}^{t(i)}_{cc} - \mathbf{X}^{s(i)}_{cc}\mathbf{W}_{cc}^{(i)} \Vert^{2}_{F} $
\Statex \hspace{0.5cm} \quad \qquad \qquad \qquad $ + \alpha \Vert \mathbf{W}^{(i)}_{cc}\Vert_{1}~(i = \{1,\cdots,C_{cc}\})$;
\State \hspace{0.5cm}${\mathbf{W}_{fc}^{(i)}}^{(k+1)} = \arg\min_{\mathbf{W}_{cc}^{(i)}} \Vert \mathbf{X}^{t(i)}_{cc} - \mathbf{X}^{s(i)}_{fc}\mathbf{W}_{fc}^{(i)} \Vert^{2}_{F}$ 
\Statex \hspace{0.5cm} \quad \qquad \qquad \qquad $ + \alpha \Vert \mathbf{W}^{(i)}_{fc}\Vert_{1}~(i = \{1,\cdots,C\})$;
\State Update $\{{\mathbf y_{i}^{t}}^{(k+1)} = g(f_3(f_2(f_1(\mathcal{X}_i^t))))\}_{i=1}^{N_t} $;
\State $k = k + 1$;
\EndFor
\end{algorithmic}
\label{al1}
\end{algorithm}

\subsubsection{Optimization Problem}

Similar to DIDAN, we combine the DR and LIDA losses in Eqs.~(\ref{eqn:nb4}) and (\ref{eqn:nb5}) into the objective function of the proposed LIDAN. Accordingly, the corresponding optimization problem can be formulated as follows:
\begin{eqnarray}
\min_{\substack{\mathbf{y}^{t}_i, \theta_{f_1}, \theta_{f_2}, \theta_{f_3}, \theta_{g}, \\ \mathbf{W}_m, \mathbf{W}_{cc}^{(i)}, \mathbf{W}_{fc}^{(i)}}} \mathcal{L}_{dr} + \lambda\mathcal{L}_{lida},
\label{eqn:nb9}
\end{eqnarray}
where $\lambda$ is the trade-off parameter to control the balance between DR and Layer-Adated losses, $\theta_{f_1}$, $\theta_{f_2}$, $\theta_{f_3}$, and $\theta_{g}$ are the parameters associated with the operations performed by the layers, $f_1$, $f_2$, $f_3$, and $g$, and $\mathbf{y}^{t}_i$ corresponds to the pseudo emotion label of $i^{th}$ target speech sample, respectively. 

It is worth noting that the emotion information of target speech samples are unlabeled in the task of cross-corpus SER. Therefore, we use their pseudo emotion labels $\{\mathbf{y}^{t}_i\}$ to serve as the model parameters to learn in the optimization of the proposed LIDAN.

\subsection{Optimization of LIDAN}

The optimization problem of LIDAN can be efficiently solved by using the alternated direction method, i.e., alternatively updating the model parameters until convergence. Specifically, we initialize pseudo emotion labels, $\mathbf{y}^{t}_i$'s, for target speech samples and then repeat the following three steps: 
\begin{enumerate}
\item Fix the target pseudo emotion labels $\{\mathbf{y}^{t}_i\}$, and update the network parameters $\{\theta_{f_1}, \theta_{f_2}, \theta_{f_3}, \theta_g\}$ and target sample reconstruction coefficient matrices $\{\mathbf{W}_m, \mathbf{W}_{cc}^{(i)}, \mathbf{W}_{fc}^{(i)}\}$.
\item Fix $\{\theta_{f_1}, \theta_{f_2}, \theta_{f_3}, \theta_g\}$ and $\{\mathbf{W}_m, \mathbf{W}_{cc}^{(i)}, \mathbf{W}_{fc}^{(i)}\}$, and update $\{\mathbf{y}^{t}_i\}$.
\item Check convergence or Reach maximal iterations.
\end{enumerate}

The complete updating procedures of training LIDAN for the issue of cross-corpus SER are given in Algorithm~\ref{al1}.

\section{Experiments}
\label{sec:exp}
\subsection{Speech Emotion Corpora and Experiment Protocol}

In this section, we conduct extensive cross-corpus SER experiments to evaluate the proposed LIDAN model. To this end, we employ three widely-used speech emotion corpora: EmoDB~\cite{burkhardt2005database}, eNTERFACE~\cite{martin2006enterface}, and CASIA~\cite{zhang2008design}. EmoDB is a German speech emotion corpus that consists of 535 audio samples from 10 professional German actors/actresses (five females and five males). Each speech sample was recorded when the corresponding actor or actress spoke a sentence expressing one of seven emotional states: \textit{Anger}, \textit{Boredom}, \textit{Disgust}, \textit{Fear}, \textit{Happiness}, \textit{Neutral}, and \textit{Sadness}. eNTERFACE is an English bimodal emotion database that contains 1,257 video clips representing six basic emotions: \textit{Anger}, \textit{Disgust}, \textit{Fear}, \textit{Happiness}, \textit{Sadness}, and \textit{Surprise}. For our experiments, we extract the audio data from eNTERFACE to design cross-corpus SER tasks. CASIA is a Chinese speech emotion corpus comprising four speakers, including two males and two females. It involves the collection of 1,200 speech samples, with each speaker required to speak 300 sentences in Chinese. The corpus covers six different emotions: \textit{Anger}, \textit{Fear}, \textit{Happiness}, \textit{Neutral}, \textit{Sadness}, and \textit{Surprise}.

\begin{table*}[t!]
\centering
\renewcommand{\arraystretch}{1.3}
\caption{The detailed data of speech emotion corpora used for all the cross-corpus SER tasks.}
\begin{tabular}{|c|c|c|}
\hline
\textbf{Task ID} & \textbf{Speech Corpus (\# Samples Belonging to Each Emotion)} & \textbf{\# Total Sample} \\ \hline \hline
B$\rightarrow$E & EmoDB (Anger: 127, Fear: 69, Disgust: 46, Happiness: 71, Sadness: 62) & 375 \\
E$\rightarrow$B & eNTERFACE (Anger: 211, Fear: 211, Disgust: 211, Happiness: 208, Sadness: 211) & 1052 \\\hline
B$\rightarrow$C & EmoDB (Anger: 127, Fear: 69, Neutral: 79, Happiness: 71, Sadness: 62) & 408 \\
C$\rightarrow$B & CASIA (Anger: 200, Fear: 200, Neutral: 200, Happiness: 200, Sadness: 200) & 1000 \\ 
\hline
E$\rightarrow$C & eNTERFACE (Anger: 211, Fear: 211, Happiness: 208, Sadness: 211, Surprise: 211) & 1052 \\
C$\rightarrow$E & CASIA (Anger: 200, Fear: 200, Happiness: 200, Sadness: 200, Surprise: 200) & 1000 \\\hline 
\end{tabular}
\label{tab:nb1}
\end{table*}

By using two of these three speech emotion corpora to alternatively serve as the source and target ones, we establish six cross-corpus SER tasks denoted as B $\rightarrow$ E, E $\rightarrow$ B, B $\rightarrow$ C, C $\rightarrow$ B, E $\rightarrow$ C, and C $\rightarrow$ E, respectively. Herein, B, E, and C represent EmoDB, eNTERFACE, and CASIA corpora, respectively. The left and right corpora in the arrow correspond to the source and target corpora for their respective cross-corpus SER task. As these corpora lack completely consistent emotion labels, we only select speech samples with matching emotion labels for each task. To provide readers with a detailed understanding of these cross-corpus SER tasks, we present the statistical information of the data used in our experiments in Table~\ref{tab:nb1}. Regarding the performance metric, we choose the unweighted average recall (UAR)~\cite{schuller2010cross}, which is defined as the average accuracy across all emotion classes. Specifically, UAR is calculated as UAR $= \frac{1}{C}\sum_{i=1}^{C} \frac{N_{i}^{p}}{N_{i}^{g}}\times100$, where $C$ represents the total number of emotion classes, and $N_{i}^{p}$ and $N_{i}^{g}$ are the numbers of samples predicted as and belonging to the $i^{th}$ emotion, respectively.

\subsection{Comparison Methods and Implimentation Details}

To evaluate the effectiveness of our proposed LIDAN method in addressing the issue of cross-corpus SER, we conduct a comparative study involving recent state-of-the-art unsupervised DA methods. Specifically, we compare our method with five \textit{Subspace Learning}: transfer component analysis (TCA)~\cite{pan2010domain}, geodesic flow kernel (GFK)~\cite{gong2012geodesic}, subspace alignment (SA)~\cite{fernando2013unsupervised}, domain-adaptive subspace learning (DoSL)~\cite{liu2018unsupervised}, and joint distribution adaptive regression (JDAR)~\cite{zhang2021cross}. Additionally, we evaluate our LIDAN against six \textit{Deep Learning} approaches: deep adaptation network (DAN)~\cite{long2015learning}, joint adaptation network (JAN)~\cite{long2017deep}, deep subdomain adaptation network (DSAN)~\cite{zhu2020deep}, domain-adversarial neural network (DANN)~\cite{ajakan2014domain}, conditional domain adversarial network (CDAN)~\cite{long2018conditional}, and DIDAN (the conference version of LIDAN)~\cite{zhao2023deep}.

Since the subspace learning methods require the feature set to describe the speech signals before adaptation, we choose two widely-used speech feature sets in the experiments, i.e., the feature set provided by the INTERSPEECH 2009 Emotion Challenge (IS09)~\cite{schuller2009interspeech} and the extended Geneva minimalistic acoustic parameter set (eGeMAPS)~\cite{eyben2015geneva}. The IS09 feature set consists of 384 elements derived from 16 low-level descriptors (LLDs) such as fundamental frequency (F0), harmonics-to-noise ratio (HNR), and Mel-frequency cepstral coefficients (MFCC) and their one-order differences through 12 statistical functions, e.g., mean, maximal, and minimal values. The eGeMAPS feature set is comprised of 25 LLDs, e.g., pitch, shimmer, Alpha ratio, and rate of loudness peaks, and 10 functions yeilding 88 parameters eventually. These two feature sets can be conveniently extracted from the speech signals by using the openSMILE toolkit~\cite{eyben2010opensmile}. 

While for the deep learning comparison methods, VGG-11~\cite{simonyan2014very} is chosen as the CNN backbone and the Mel-spectrograms of speech signals are used to serve as their input. Hence, we subsequently resize the images of Mel-spectrograms at a size of $224 \times 224$ pixels. It is also noted that since the label information is entirely unavailable in the tasks of cross-corpus SER, in the experiments we follow the tradition of unsupervised DA evaluation and report their best results by searching the trade-off parameters from a given interval. Linear support vector machine (SVM)~\cite{chang2011libsvm} is chosen with the parameter $C=1$ for all the subspace learning methods without classification ability including TCA, GFK, and SA. In addition, we also directly use the SVM and VGG-11 to respectively conduct all the cross-corpus SER experiments as the baselines. In summary, the trade-off parameters for all the unsupervised DA methods are set in the experiments as follows:

\subsubsection{TCA, GFK, and SA}

These three methods aim to seek a $d$-dimensional subspace to relieve the feature distribution mismatch between the source and target speech samples. Hence, in the experiments, $d$ is fixed at the one selected from a parameter interval, $[5:5:d_{max}]$, where $d_{max}$ is the element number of the feature set used in the experiments.

\subsubsection{DoSL and JDAR}

In these three methods, two trade-off parameters need to be set, i.e., $\lambda$ and $\mu$, respectively controlling the balance between the sparsity and feature distribution elimination terms and the original regression loss function. $\lambda$ and $\mu$ are set by searching from $[5:5:200]$ throughout all the experiments.

\subsubsection{DAN, JAN, DSAN, DANN, and CDAN}

One trade-off parameter $\lambda$ exists in the objective function to control the balance between the oringinal loss function and the domain adaptation term. In the experiments, we search it from an elaborate parameter interval $[0.0001:0.0001:0.001, 0.002:0.001:0.01, 0.02:0.01:0.1, 0.2:0.1:1, 2, 5, 10, 100]$.

\subsubsection{DIDAN and LIDAN}

Besides $\lambda$, DIDAN and LIDAN also introduce a new trade-off parameter $\alpha$ to control the sparsity of the target sample reconstruction coefficient matrix. In our experiments, the parameter ranges for both $\lambda$ and $\alpha$ are set similar to other deep learning comparison methods. Specifically, we use the intervals $[0.0001:0.0001:0.001, 0.002:0.001:0.01, 0.02:0.01:0.1, 0.2:0.1:1, 2, 5, 10, 100]$. In addition, LIDAN also requires the combination of fine-grained emotion labels into coarse-grained emotion labels to calculate the \textit{Coarse-grained Emotion Label Aware Conditional Distribution Guided IDA} ($\mathcal{L}_{lida_{(cc)}}$). In our experiments, we divide the fine emotion labels originally provided by the speech corpora into two groups based on their valence, inspired by the distribution of various emotions on recently designed arousal-valence emotion wheels~\cite{jing2019automatic,toisoul2021estimation,yang2022hybrid}. The two groups are referred to as \textit{Positive}, which includes \textit{Surprise}, \textit{Happiness}, and \textit{Neutral}, and \textit{Negative}, which includes \textit{Anger}, \textit{Disgust}, \textit{Fear}, and \textit{Sadness}. These groups serve as the coarse-grained emotion labels in our experiments.

\begin{table*}[t!]
\centering
\renewcommand{\arraystretch}{1.3}
\caption{The comparison results in term of UAR between the proposed LIDAN and subspace learning-based DA methods in addressing six cross-corpus SER tasks. The best result for each task is highlighted in bold.}
\begin{tabular}{|c|c|c|cccccc|c|}
\hline
\multicolumn{2}{|c|}{\textbf{Method}} & \textbf{Alignment Type} & \textbf{B$\rightarrow$ E} & \textbf{E$\rightarrow$B} & \textbf{B$\rightarrow$C} & \textbf{C$\rightarrow$B} & \textbf{E$\rightarrow$C} & \textbf{C$\rightarrow$E} & \textbf{Average}\\
\hline \hline
\multirow{6}{*}{\makecell[c]{Subspace Learning \\ (IS09 Feature Set)}} & SVM & - & ~~28.93~~ & ~~23.58~~ & ~~29.60~~ & ~~35.01~~ & ~~26.10~~ & ~~25.14~~ & ~28.06~~\\
& TCA & Explicit & 30.52 & 44.03 & 33.40 & 45.07 & 31.10 & 32.32 & 36.07\\
& GFK & Explicit & 32.11 & 42.48 & 33.10 & 48.08 & 32.80 & 28.13 & 36.17\\
& SA & Explicit & 33.50 & 43.89 & 35.80 & 49.03 & 32.60 & 28.17 & 36.33\\
& DoSL & Explicit & 36.12 & 38.95 & 34.40 & 45.75 & 30.40 & 31.59 & 36.20\\
& JDAR & Explicit & \textbf{36.33} & 39.97 & 31.10 & 46.29 & 32.40 & 31.50 & 36.27 \\ \hline
\multirow{6}{*}{\makecell[c]{Subspace Learning \\ (eGeMAPS Feature Set)}} & SVM & - & 26.65 & 32.58 & 33.50 & 51.84 & 36.40 & 34.79 & 35.96\\
& TCA & Explicit & 28.55 & 33.29 & \textbf{42.90} & 49.59 & \textbf{40.80} & 34.95 & 38.35 \\
& GFK & Explicit & 29.42 & 33.86 & 36.20 & 50.13 & 38.50 & 34.00 & 37.02\\
& SA & Explicit & 32.11 & 36.78 & 34.80 & 52.00 & 37.00 & 35.32 & 38.00 \\
& DoSL & Explicit & 30.52 & 40.17 & 34.40 & 52.05 & 38.30 & 32.92 & 38.06 \\
& JDAR & Explicit & 31.00 & 43.48 & 38.60 & 55.11 & 38.30 & 32.89 & 39.90 \\ \hline\hline
Deep Learning & LIDAN (Ours) & Implicit & 34.44 & \textbf{47.04} & 39.00 & \textbf{58.06} & 33.88 & \textbf{34.97} & \textbf{41.23} \\\hline
\end{tabular}
\label{tab:nb2}
\end{table*}

\begin{table*}[t!]
\centering
\renewcommand{\arraystretch}{1.3}
\caption{The comparison results in term of UAR between the proposed LIDAN and deep learning-based DA methods in addressing six cross-corpus SER tasks. The best result for each task is highlighted in bold.}
\begin{tabular}{|c|c|c|cccccc|c|}
\hline
\multicolumn{2}{|c|}{\textbf{Method}} & \textbf{Alignment Type} & \textbf{B$\rightarrow$ E} & \textbf{E$\rightarrow$B} & \textbf{B$\rightarrow$C} & \textbf{C$\rightarrow$B} & \textbf{E$\rightarrow$C} & \textbf{C$\rightarrow$E} & \textbf{Average}\\
\hline \hline
\multirow{8}{*}{Deep Learning}& VGG-11 & - & ~~27.08~~ & ~~34.83~~ & ~~34.80~~ & ~~51.31~~ & ~~26.90~~ & ~~26.02~~ & ~~33.49~~ \\
& DAN& Explicit & 33.58 & 43.50 & 36.30 & 56.72 & 29.30 & 32.17& 38.60 \\
& JAN & Explicit& 35.23 & 47.29 & 37.00 & 57.51 & 31.00 & 32.21& 40.04 \\
& DSAN& Explicit & 31.82 & \textbf{47.58} & 35.80 & 56.50 & 29.00 & 31.25 & 38.66 \\
& DANN & Implicit& 32.56 & 46.06 & 36.40 & 57.67 & 30.50 & 33.77 & 39.49 \\
& CDAN& Implicit & 31.62 & 46.12 & 35.40 & 57.60 & 30.30 & 33.49& 39.09 \\ \cline{2-10}
& DIDAN (Ours) & Implicit & 33.05& 47.11 & 38.90 & 56.22 & 31.10 & 34.06 & 40.07 \\
& LIDAN (Ours) & Implicit & \textbf{34.44} & 47.04 & \textbf{39.00} & \textbf{58.06} & \textbf{33.88} & \textbf{34.97} & \textbf{41.23} \\\hline
\end{tabular}
\label{tab:nb3}
\end{table*}

\subsection{Comparison with State-of-the-art Subspace Learning-Based Unsupervised DA Methods}

We begin by comparing the performance of our LIDAN method with recent state-of-the-art subspace learning-based unsupervised DA methods in coping with the tasks of cross-corpus SER. The results are presented in Table~\ref{tab:nb2}. From the table, several interesting observations and conclusions can be drawn.

Firstly, our proposed LIDAN method achieved the highest average UAR among all the methods, reaching 41.23\%. LIDAN also outperformed all the comparison methods in three out of the six cross-corpus SER tasks: E $\rightarrow$ B (47.04\%), C $\rightarrow$ B (58.06\%), and C $\rightarrow$ E (34.97\%). While LIDAN did not achieve the best performance in the remaining three tasks, it still demonstrated high competitiveness against the best-performing methods. For example, in the task, B $\rightarrow$ E, LIDAN achieved a UAR of 34.44\% compared to 36.33\% obtained by JDAR + IS09. These observations demonstrate the effectiveness and overall superior performance of our proposed LIDAN in addressing the challenge of cross-corpus SER.

Secondly, it is worth noting that subspace learning methods utilizing the eGeMAPS feature set to describe speech signals attained promisingly better results than our LIDAN in coping with the cross-corpus SER task, E $\rightarrow$ C. In this case, all the subspace learning methods, including the baseline SVM without any DA, outperformed LIDAN with an increase in UAR of at least 2.52\%. This could be attributed to the fact that the hand-crafted speech feature set used in subspace learning methods, namely eGeMAPS, is more emotion-discriminative and corpus-invariant compared to the deep features learned directly from speech Mel-spectrum images using the VGG-11 backbone in LIDAN. Further comparison results of these DA methods in the other task involving the same speech corpora, i.e., C $\rightarrow$ E, confirm that these DA methods generally exhibit competitive performance against our LIDAN, thereby supporting our explanations and analysis.

Last but not least, it is interesting to observe that all subspace learning methods utilizing the eGeMAPS feature set achieved substantially higher UAR than those using the IS09 feature set in coping with cross-corpus SER tasks. This indicates that each subspace learning-based DA method often exhibits different performances, with a considerable gap, when employing different speech feature sets to address cross-corpus SER tasks. For example, in our experiments, the baseline method (SVM without DA) achieved an average UAR increase from 28.06\% to 35.96\% when transitioning the feature set from IS09 to eGeMAPS. Moreover, SVM, TCA, and JDAR also experienced promising increases of over 2\% in terms of average UAR when replacing IS09 with eGeMAPS. These observations highlight the importance of selecting suitable feature sets to describe speech signals when employing subspace learning methods to tackle the issue of cross-corpus SER.

\subsection{Comparison with State-of-the-art Deep Learning-Based Unsupervised DA Methods}
\label{sec:dl}

In addition to the comparison with subspace learning methods, we also compared our LIDAN with recent state-of-the-art deep learning-based unsupervised DA methods in Table~\ref{tab:nb3}. It can be observed that LIDAN achieved the highest average UAR among all the deep learning methods for recognizing emotions from speech signals across different corpora, surpassing them in five out of six cross-corpus SER tasks. Even for the task of E $\rightarrow$ B, there is only a negligible difference in performance between LIDAN (47.04\%) and the best-performing comparison method, DSAN (47.58\%). These results further demonstrate the effectiveness and superior performance of LIDAN compared to recent state-of-the-art deep learning methods, consistent with the comparisons made with the aforementioned subspace learning methods.

Furthermore, we also highlighted the alignment types used in all deep learning methods. It can be clearly observed from Table~\ref{tab:nb3} that both the proposed LIDAN and its origin, DIDAN, achieved more promising performance than all the explicit methods that adopt MMD and its variants to measure the feature distribution gap between the source and target speech samples. Moreover, in contrast to current widely-used adversarial learning-based implicit methods, our LIDAN and DIDAN exhibit superior performance as well, resulting in a remarkable UAR increase. These comparisons validate the effectiveness and superiority of the proposed implicit feature distribution alignment idea, namely target sample reconstruction, utilized in LIDAN and DIDAN in bridging the feature distribution gap across the speech emotion corpora.

Additionally, we conducted a comparison between LIDAN and DIDAN. Table~\ref{tab:nb3} demonstrates that LIDAN achieved an average UAR of 41.23\% across all six cross-corpus SER tasks, while DIDAN only achieved 40.07\%. In particular, LIDAN outperformed DIDAN in four out of the six SER tasks, showcasing a satisfactory performance increase. Moreover, even in the remaining two tasks, the performance of LIDAN only slightly dropped when compared to DIDAN. Note that the major difference between LIDAN and DIDAN can be observed in Eq.~(\ref{eqn:nb3}) and (\ref{eqn:nb9}), where LIDAN extended the original IDA utilized in DIDAN to a layer-adapted version, LIDA. These comparative findings strongly support the effectiveness and feasibility of extending IDA to LIDA as evidence of its enhanced capability in addressing the challenge of cross-corpus SER compared to the original IDA.

%
%

\subsection{Going Deeper into Layer-Adapted IDA in LIDAN}

As mentioned previously, the main contribution of our LIDAN method lies in the introduction of LIDA, which builds upon the IDA utilized in DIDAN. LIDA aims to further enhance the robustness of DIDAN to corpus variance by incorporating emotion class aware conditional IDA terms alongside the original IDA. These terms are strategically incorporated into the fully connected layers, with a progressive integration from shallow to deep layers. In this subsection, we select three representative cross-corpus SER tasks, namely, B $\rightarrow$ E, E $\rightarrow$ B, and B $\rightarrow$ C, to conduct additional experiments. The objective is to address the three key questions regarding LIDA, allowing readers to gain a deeper understanding of this well-designed module in LIDAN.

\subsubsection{Are All the Modified Terms Necessary for Layer-Adapted IDA of LIDAN?}

The LIDA module, as illustrated in Eq.~(\ref{eqn:nb5}), consists of three modified IDA terms: $\mathcal{L}_{ida_m}$, $\mathcal{L}_{ida_{(cc)}}$, and $\mathcal{L}_{ida_{(fc)}}$. Among them, $\mathcal{L}_{ida_m}$ follows a similar formulation to the original IDA, while $\mathcal{L}_{ida_{(cc)}}$ and $\mathcal{L}_{ida_{(fc)}}$ are newly-introduced modifications that take conditional distributions into account. It is natural to question whether all three modified IDA terms are necessary for improving the performance of LIDAN in tackling the tasks of cross-corpus SER. To address this concern, we conducted ablation experiments on three representative cross-corpus SER tasks. Specifically, we systematically removed one or more terms from the LIDA module and trained LIDAN accordingly. The experimental results, presented in Table~\ref{tab:nb4}, clearly demonstrate that LIDAN, utilizing all three modified IDA terms in the LIDA module, achieved the best performance in terms of UAR for all the tasks. This indicates that each of the designed modified IDA terms contributes significantly to enhancing the robustness of LIDAN towards corpus variance in cross-corpus SER problems.

\begin{table}[t]
\caption{Experimental results of the ablation analysis for modified IDA terms in Layer-Adapted IDA, where the best results are highlight in bold (\%).}
\centering
\renewcommand{\arraystretch}{1.3}
\begin{tabular}{|c|c|c|ccc|}
\hline
\multicolumn{3}{|c|}{\textbf{Ablation of Modified IDA Terms}}& \multicolumn{3}{c|}{\textbf{Task ID}} \\ \hline
$\mathcal{L}_{ida_m}$ & $\mathcal{L}_{ida_{(cc)}}$ & $\mathcal{L}_{ida_{(fc)}}$ & \textbf{B $\rightarrow$ E}&\textbf{E $\rightarrow$ B}& \textbf{B $\rightarrow$ C} \\ \hline \hline
\xmark & \xmark & \xmark & 33.58 & 43.50 & 36.30 \\
\cmark & \xmark & \xmark & 31.22 & 42.69 & 35.90 \\
\xmark & \cmark & \xmark & 31.35 & 43.56 & 35.60 \\
\xmark & \xmark & \cmark & 31.20 & 41.18 & 35.80 \\
\xmark & \cmark & \cmark & 32.16 & 42.83 & 36.30 \\
\cmark & \xmark & \cmark & 30.68 & 41.00 & 37.50 \\
\cmark & \cmark & \xmark & 31.86 & 41.01 & 36.80 \\\hline 
\cmark & \cmark & \cmark & \textbf{34.44} & \textbf{47.07} & \textbf{39.00} \\ \hline 
\end{tabular}
\label{tab:nb4}
\end{table}

\subsubsection{How Should the Modified IDA Terms Be Arranged within Different Layers in LIDAN?}

From Eqs.~(\ref{eqn:nb6}), (\ref{eqn:nb7}), and (\ref{eqn:nb8}), three modified IDA terms: $\mathcal{L}_{ida_m}$, $\mathcal{L}_{ida_{(cc)}}$, and $\mathcal{L}_{ida_{(fc)}}$ in \textit{Layer-Adapted IDA} module make use of varying levels of emotion labels, including no, coarse-grained, and fine-grained labels. Consequently, the arrangement of these terms in LIDAN follows a progression from shallow to deep fully connected layers, where the features learned by these layers exhibit increasingly discriminative abilities for emotions with respect to the layer depth. This raises the question of whether aligning the arrangement of these terms with the emotion-discriminative abilities of layers, based on their requirement of emotion label types, is a suitable choice. To investigate this point, we conducted experiments in LIDAN by exploring all possible permutations for these three terms. The experimental results are presented in Table~\ref{tab:nb5}. From the table, it is evident that the current arrangement of these modified IDA terms in LIDAN achieved the highest UAR performance among all possible permutations. This demonstrates the effectiveness and suitability of our designed arrangement for the LIDA module in LIDAN.

\begin{table}[t!]
\caption{Experimental results of all possible permutations of modified IDA terms in Layer-Adapted IDA, where the best results are highlight in bold.}
\centering
\renewcommand{\arraystretch}{1.3}
\begin{tabular}{|c|c|c|ccc|}
\hline
\multicolumn{3}{|c|}{\textbf{Rankings of Modified IDA Terms}}& \multicolumn{3}{c|}{\textbf{Task ID}} \\ \hline
~~$\mathcal{L}_{ida_m}$~~ & ~$\mathcal{L}_{ida_{(cc)}}$~ & $\mathcal{L}_{ida_{(fc)}}$ & \textbf{B $\rightarrow$ E}&\textbf{E $\rightarrow$ B}& \textbf{B $\rightarrow$ C} \\ \hline \hline
1 & 3 & 2 & 32.55 & 43.25 & 37.40 \\ 
2 & 1 & 3 & 32.95 & 42.86 & 38.70 \\ 
2 & 3 & 1 & 32.26 & 42.67 & 36.80 \\ 
3 & 1 & 2 & 33.36 & 42.60 & 37.60 \\ 
3 & 2 & 1 & 31.84 & 41.26 & 36.80 \\ \hline 
\textbf{1} & \textbf{2} & \textbf{3} & \textbf{34.44} & \textbf{47.07} & \textbf{39.00} \\ \hline 
\end{tabular}
\label{tab:nb5}
\end{table}

\begin{figure*}[t!]
\centering
\includegraphics[width=0.8\textwidth]{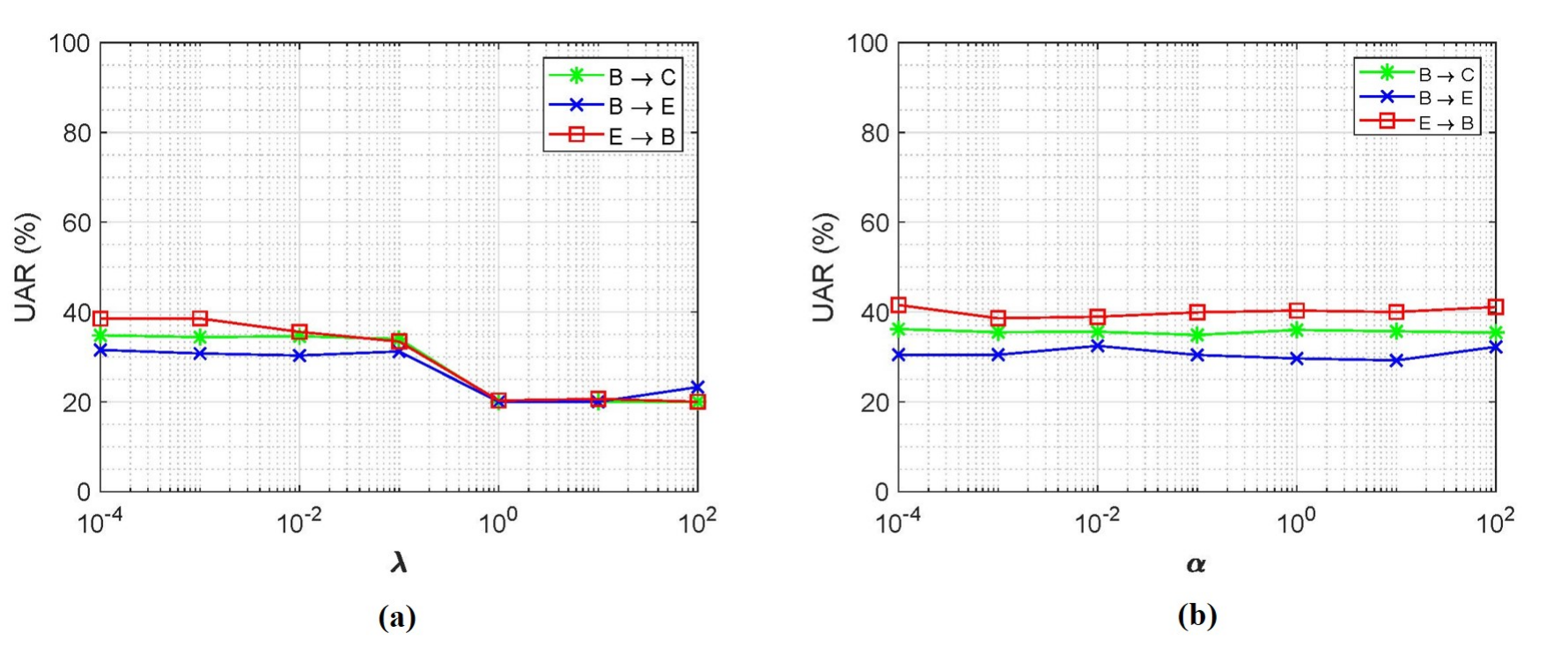}
\caption{Results of trade-off parameter sensitivity analysis experiments for the proposed LIDAN method. (a) illustrates the outcomes obtained by varying the value of $\lambda$, whereas (b) depicts the results obtained by varying the value of $\alpha$.}
\label{fig:nb2}
\end{figure*}

\subsubsection{Is "Sparsely" Necessary for Target Sample Reconstruction in LIDAN?}

Similar to IDA module in DIDAN, the LIDA module in LIDAN utilizes the approach of target sample reconstruction to address the feature distribution mismatch between the source and target speech signals. It is worth noting that the LIDA module incorporates a sparse regularization term, as shown in Eqs.~(\ref{eqn:nb6}), (\ref{eqn:nb7}), and (\ref{eqn:nb8}), which encourages LIDAN to select a few source samples for reconstructing the target samples. This raises another question of whether the sparse regularization term is necessary for achieving target sample reconstruction. In order to explore this further, we conducted cross-corpus SER experiments by excluding the sparse regularization term from the total loss function of both DIDAN and LIDAN. The experimental results, shown in Table~\ref{tab:nb6}, demonstrate that the inclusion of the sparse regularization term leads to a significant performance improvement for both models compared to their non-sparse versions. These findings confirm the effectiveness of incorporating the sparse regularization term to guide the learning of the target reconstruction coefficient matrix in DIDAN and LIDAN.

\begin{table}[t!]
\centering
\footnotesize
\caption{Experimental results of the necessity analysis for sparsity regularization term in target sample reconstruction, where the best results in the cases of DIDAN and LIDAN are highlighted in bold, respectively.}
\renewcommand{\arraystretch}{1.3}
\begin{tabular}{|l|ccc|}
\hline
\textbf{Method} & \textbf{B$\rightarrow$ E} & \textbf{E$\rightarrow$B} & \textbf{B$\rightarrow$C} \\
\hline\hline
VGG-11 & 33.58 & 43.50 & 36.30 \\\hline
DIDAN w nonSparse IDA & 31.83 & 41.69 & 37.70\\
DIDAN w Sparse IDA & \textbf{33.05} & \textbf{47.11} & \textbf{38.90}\\\hline
LIDAN w nonSparse Layer-Adapted IDA & 32.48 & 42.32 & 34.50\\
LIDAN w Sparse Layer-Adapted IDA & \textbf{34.44} & \textbf{47.07} & \textbf{39.00} \\
\hline
\end{tabular}
\label{tab:nb6}
\end{table}

\subsection{Sensitivity Analysis of Trade-off Parameters in LIDAN}

The proposed LIDAN method involves two crucial trade-off parameters: $\lambda$ and $\alpha$. The selection of these parameters directly impacts the performance of LIDAN. In our experiments, we utilized fixed values for these parameters, which prompted concerns regarding the sensitivity of LIDAN's performance to changes in their values. To investigate this, we conducted experiments on three previously used cross-corpus SER tasks: B $\rightarrow$ E, E $\rightarrow$ B, and B $\rightarrow$ C. In these experiments, we alternatively held one parameter constant while varying the other. The intervals for varied $\lambda$ and $\alpha$ were set as $[0.0001, 0.01, 0.1, 1, 10, 100]$. The constant values for $\lambda$ and $\alpha$ matched those used in the experiments described in Section~\ref{sec:dl}. Fig.~\ref{fig:nb2} presents the experimental results. As depicted in the figure, it is clear that the performance shows less variation with respect to changes in $\lambda$ and $\alpha$ for all three experiments. This suggests that our proposed LIDAN method is relatively insensitive to changes in its trade-off parameters.

\section{Conclusion}
\label{sec:conclusion}

In this paper, we focus on the research of cross-corpus SER from the perspective of implicit distribution alignment. We have proposed a novel deep learning-based unsupervised DA method called LIDAN by building upon our previous work at ICASSP, DIDAN. The key improvement of LIDAN compared to DIDAN lies in the design of an extended IDA called LIDA. This allows LIDAN to implicitly and effectively align feature distributions of source and target speech samples across multiple layers, considering the emotion-discriminative capability of each layer's features. As a result, LIDAN is more advantageous in learning both emotion-discriminative and corpus-invariant features for cross-corpus SER, outperforming DIDAN. We evaluated the performance of LIDAN through extensive cross-corpus SER experiments on the EmoDB, eNTERFACE, and CASIA corpora. The experimental results demonstrate that LIDAN exceeds recent state-of-the-art methods based on unsupervised DA using subspace learning and deep learning techniques in tackling cross-corpus SER tasks. Notably, our work, including DIDAN and LIDAN, is the first to highlight the distinction between implicit and explicit approaches in cross-corpus SER methods.



%

\bibliographystyle{IEEEtran}
\bibliography{TCSS2023}

\begin{thebibliography}{10}
\providecommand{\url}[1]{#1}
\csname url@samestyle\endcsname
\providecommand{\newblock}{\relax}
\providecommand{\bibinfo}[2]{#2}
\providecommand{\BIBentrySTDinterwordspacing}{\spaceskip=0pt\relax}
\providecommand{\BIBentryALTinterwordstretchfactor}{4}
\providecommand{\BIBentryALTinterwordspacing}{\spaceskip=\fontdimen2\font plus
\BIBentryALTinterwordstretchfactor\fontdimen3\font minus
  \fontdimen4\font\relax}
\providecommand{\BIBforeignlanguage}[2]{{%
\expandafter\ifx\csname l@#1\endcsname\relax
\typeout{** WARNING: IEEEtran.bst: No hyphenation pattern has been}%
\typeout{** loaded for the language `#1'. Using the pattern for}%
\typeout{** the default language instead.}%
\else
\language=\csname l@#1\endcsname
\fi
#2}}
\providecommand{\BIBdecl}{\relax}
\BIBdecl

\bibitem{el2011survey}
M.~El~Ayadi, M.~S. Kamel, and F.~Karray, ``Survey on speech emotion
  recognition: Features, classification schemes, and databases,'' \emph{Pattern
  Recognition}, vol.~44, no.~3, pp. 572--587, 2011.

\bibitem{akccay2020speech}
M.~B. Ak{\c{c}}ay and K.~O{\u{g}}uz, ``Speech emotion recognition: Emotional
  models, databases, features, preprocessing methods, supporting modalities,
  and classifiers,'' \emph{Speech Communication}, vol. 116, pp. 56--76, 2020.

\bibitem{williams1972emotions}
C.~E. Williams and K.~N. Stevens, ``Emotions and speech: Some acoustical
  correlates,'' \emph{The journal of the acoustical society of America},
  vol.~52, no.~4B, pp. 1238--1250, 1972.

\bibitem{schuller2003hidden}
B.~Schuller, G.~Rigoll, and M.~Lang, ``Hidden markov model-based speech emotion
  recognition,'' in \emph{2003 IEEE International Conference on Acoustics,
  Speech, and Signal Processing, 2003. Proceedings.(ICASSP'03).}, vol.~2.\hskip
  1em plus 0.5em minus 0.4em\relax Ieee, 2003, pp. II--1.

\bibitem{wu2011automatic}
S.~Wu, T.~H. Falk, and W.-Y. Chan, ``Automatic speech emotion recognition using
  modulation spectral features,'' \emph{Speech communication}, vol.~53, no.~5,
  pp. 768--785, 2011.

\bibitem{huang2014speech}
Z.~Huang, M.~Dong, Q.~Mao, and Y.~Zhan, ``Speech emotion recognition using
  cnn,'' in \emph{Proceedings of the 22nd ACM international conference on
  Multimedia}, 2014, pp. 801--804.

\bibitem{zhang2019spontaneous}
S.~Zhang, X.~Zhao, and Q.~Tian, ``Spontaneous speech emotion recognition using
  multiscale deep convolutional lstm,'' \emph{IEEE Transactions on Affective
  Computing}, vol.~13, no.~2, pp. 680--688, 2019.

\bibitem{issa2020speech}
D.~Issa, M.~F. Demirci, and A.~Yazici, ``Speech emotion recognition with deep
  convolutional neural networks,'' \emph{Biomedical Signal Processing and
  Control}, vol.~59, p. 101894, 2020.

\bibitem{lu2022domain}
C.~Lu, Y.~Zong, W.~Zheng, Y.~Li, C.~Tang, and B.~W. Schuller, ``Domain
  invariant feature learning for speaker-independent speech emotion
  recognition,'' \emph{IEEE/ACM Transactions on Audio, Speech, and Language
  Processing}, vol.~30, pp. 2217--2230, 2022.

\bibitem{lu2022speech}
C.~Lu, W.~Zheng, H.~Lian, Y.~Zong, C.~Tang, S.~Li, and Y.~Zhao, ``Speech
  emotion recognition via an attentive time-frequency neural network,''
  \emph{IEEE Transactions on Computational Social Systems}, 2022.

\bibitem{schuller2010cross}
B.~Schuller, B.~Vlasenko, F.~Eyben, M.~W{\"o}llmer, A.~Stuhlsatz, A.~Wendemuth,
  and G.~Rigoll, ``Cross-corpus acoustic emotion recognition: Variances and
  strategies,'' \emph{IEEE Transactions on Affective Computing}, vol.~1, no.~2,
  pp. 119--131, 2010.

\bibitem{wang2018deep}
M.~Wang and W.~Deng, ``Deep visual domain adaptation: A survey,''
  \emph{Neurocomputing}, vol. 312, pp. 135--153, 2018.

\bibitem{zhuang2020comprehensive}
F.~Zhuang, Z.~Qi, K.~Duan, D.~Xi, Y.~Zhu, H.~Zhu, H.~Xiong, and Q.~He, ``A
  comprehensive survey on transfer learning,'' \emph{Proceedings of the IEEE},
  vol. 109, no.~1, pp. 43--76, 2020.

\bibitem{hassan2013acoustic}
A.~Hassan, R.~Damper, and M.~Niranjan, ``On acoustic emotion recognition:
  compensating for covariate shift,'' \emph{IEEE Transactions on Audio, Speech,
  and Language Processing}, vol.~21, no.~7, pp. 1458--1468, 2013.

\bibitem{gretton2009covariate}
A.~Gretton, A.~Smola, J.~Huang, M.~Schmittfull, K.~Borgwardt, B.~Sch{\"o}lkopf
  \emph{et~al.}, ``Covariate shift by kernel mean matching,'' \emph{Dataset
  shift in machine learning}, vol.~3, no.~4, p.~5, 2009.

\bibitem{borgwardt2006integrating}
K.~M. Borgwardt, A.~Gretton, M.~J. Rasch, H.-P. Kriegel, B.~Sch{\"o}lkopf, and
  A.~J. Smola, ``Integrating structured biological data by kernel maximum mean
  discrepancy,'' \emph{Bioinformatics}, vol.~22, no.~14, pp. e49--e57, 2006.

\bibitem{zhang2021cross}
J.~Zhang, L.~Jiang, Y.~Zong, W.~Zheng, and L.~Zhao, ``Cross-corpus speech
  emotion recognition using joint distribution adaptive regression,'' in
  \emph{ICASSP 2021-2021 IEEE International Conference on Acoustics, Speech and
  Signal Processing (ICASSP)}.\hskip 1em plus 0.5em minus 0.4em\relax IEEE,
  2021, pp. 3790--3794.

\bibitem{zhao2022deep}
Y.~Zhao, J.~Wang, R.~Ye, Y.~Zong, W.~Zheng, and L.~Zhao, ``Deep transductive
  transfer regression network for cross-corpus speech emotion recognition,''
  \emph{Proceedings of the INTERSPEECH, Incheon, Korea}, pp. 18--22, 2022.

\bibitem{abdelwahab2018domain}
M.~Abdelwahab and C.~Busso, ``Domain adversarial for acoustic emotion
  recognition,'' \emph{IEEE/ACM Transactions on Audio, Speech, and Language
  Processing}, vol.~26, no.~12, pp. 2423--2435, 2018.

\bibitem{gideon2021improving}
J.~Gideon, M.~G. McInnis, and E.~M. Provost, ``Improving cross-corpus speech
  emotion recognition with adversarial discriminative domain generalization
  (addog),'' \emph{IEEE Transactions on Affective Computing}, vol.~12, no.~4,
  pp. 1055--1068, 2021.

\bibitem{latif2022self}
S.~Latif, R.~Rana, S.~Khalifa, R.~Jurdak, and B.~W. Schuller, ``Self supervised
  adversarial domain adaptation for cross-corpus and cross-language speech
  emotion recognition,'' \emph{IEEE Transactions on Affective Computing}, 2022.

\bibitem{gao2023adversarial}
Y.~Gao, L.~Wang, J.~Liu, J.~Dang, and S.~Okada, ``Adversarial domain
  generalized transformer for cross-corpus speech emotion recognition,''
  \emph{IEEE Transactions on Affective Computing}, 2023.

\bibitem{zhao2023deep}
Y.~Zhao, J.~Wang, Y.~Zong, W.~Zheng, H.~Lian, and L.~Zhao, ``Deep implicit
  distribution alignment networks for cross-corpus speech emotion
  recognition,'' in \emph{ICASSP 2023-2023 IEEE International Conference on
  Acoustics, Speech and Signal Processing (ICASSP)}.\hskip 1em plus 0.5em minus
  0.4em\relax IEEE, 2023, pp. 1--5.

\bibitem{long2013transfer}
M.~Long, J.~Wang, G.~Ding, J.~Sun, and P.~S. Yu, ``Transfer feature learning
  with joint distribution adaptation,'' in \emph{Proceedings of the IEEE
  international conference on computer vision}, 2013, pp. 2200--2207.

\bibitem{burkhardt2005database}
F.~Burkhardt, A.~Paeschke, M.~Rolfes, W.~F. Sendlmeier, B.~Weiss \emph{et~al.},
  ``A database of german emotional speech.'' in \emph{Interspeech}, vol.~5,
  2005, pp. 1517--1520.

\bibitem{martin2006enterface}
O.~Martin, I.~Kotsia, B.~Macq, and I.~Pitas, ``The enterface'05 audio-visual
  emotion database,'' in \emph{22nd International Conference on Data
  Engineering Workshops (ICDEW'06)}.\hskip 1em plus 0.5em minus 0.4em\relax
  IEEE, 2006, pp. 8--8.

\bibitem{zhang2008design}
J.~Zhang and H.~Jia, ``Design of speech corpus for mandarin text to speech,''
  in \emph{The blizzard challenge 2008 workshop}, 2008.

\bibitem{pan2010domain}
S.~J. Pan, I.~W. Tsang, J.~T. Kwok, and Q.~Yang, ``Domain adaptation via
  transfer component analysis,'' \emph{IEEE transactions on neural networks},
  vol.~22, no.~2, pp. 199--210, 2010.

\bibitem{gong2012geodesic}
B.~Gong, Y.~Shi, F.~Sha, and K.~Grauman, ``Geodesic flow kernel for
  unsupervised domain adaptation,'' in \emph{2012 IEEE conference on computer
  vision and pattern recognition}.\hskip 1em plus 0.5em minus 0.4em\relax IEEE,
  2012, pp. 2066--2073.

\bibitem{fernando2013unsupervised}
B.~Fernando, A.~Habrard, M.~Sebban, and T.~Tuytelaars, ``Unsupervised visual
  domain adaptation using subspace alignment,'' in \emph{Proceedings of the
  IEEE international conference on computer vision}, 2013, pp. 2960--2967.

\bibitem{liu2018unsupervised}
N.~Liu, Y.~Zong, B.~Zhang, L.~Liu, J.~Chen, G.~Zhao, and J.~Zhu, ``Unsupervised
  cross-corpus speech emotion recognition using domain-adaptive subspace
  learning,'' in \emph{2018 IEEE International Conference on Acoustics, Speech
  and Signal Processing (ICASSP)}.\hskip 1em plus 0.5em minus 0.4em\relax IEEE,
  2018, pp. 5144--5148.

\bibitem{long2015learning}
M.~Long, Y.~Cao, J.~Wang, and M.~Jordan, ``Learning transferable features with
  deep adaptation networks,'' in \emph{International conference on machine
  learning}.\hskip 1em plus 0.5em minus 0.4em\relax PMLR, 2015, pp. 97--105.

\bibitem{long2017deep}
M.~Long, H.~Zhu, J.~Wang, and M.~I. Jordan, ``Deep transfer learning with joint
  adaptation networks,'' in \emph{International conference on machine
  learning}.\hskip 1em plus 0.5em minus 0.4em\relax PMLR, 2017, pp. 2208--2217.

\bibitem{zhu2020deep}
Y.~Zhu, F.~Zhuang, J.~Wang, G.~Ke, J.~Chen, J.~Bian, H.~Xiong, and Q.~He,
  ``Deep subdomain adaptation network for image classification,'' \emph{IEEE
  transactions on neural networks and learning systems}, vol.~32, no.~4, pp.
  1713--1722, 2020.

\bibitem{ajakan2014domain}
H.~Ajakan, P.~Germain, H.~Larochelle, F.~Laviolette, and M.~Marchand,
  ``Domain-adversarial neural networks,'' \emph{arXiv preprint
  arXiv:1412.4446}, 2014.

\bibitem{long2018conditional}
M.~Long, Z.~Cao, J.~Wang, and M.~I. Jordan, ``Conditional adversarial domain
  adaptation,'' \emph{Advances in neural information processing systems},
  vol.~31, 2018.

\bibitem{schuller2009interspeech}
B.~Schuller, S.~Steidl, and A.~Batliner, ``The interspeech 2009 emotion
  challenge,'' in \emph{INTERSPEECH}, 2009.

\bibitem{eyben2015geneva}
F.~Eyben, K.~R. Scherer, B.~W. Schuller, J.~Sundberg, E.~Andr{\'e}, C.~Busso,
  L.~Y. Devillers, J.~Epps, P.~Laukka, S.~S. Narayanan \emph{et~al.}, ``The
  geneva minimalistic acoustic parameter set (gemaps) for voice research and
  affective computing,'' \emph{IEEE Transactions on Affective Computing},
  vol.~7, no.~2, pp. 190--202, 2015.

\bibitem{eyben2010opensmile}
F.~Eyben, M.~W{\"o}llmer, and B.~Schuller, ``Opensmile: the munich versatile
  and fast open-source audio feature extractor,'' in \emph{Proceedings of the
  18th ACM international conference on Multimedia}, 2010, pp. 1459--1462.

\bibitem{simonyan2014very}
K.~Simonyan and A.~Zisserman, ``Very deep convolutional networks for
  large-scale image recognition,'' \emph{arXiv preprint arXiv:1409.1556}, 2014.

\bibitem{chang2011libsvm}
C.-C. Chang and C.-J. Lin, ``Libsvm: a library for support vector machines,''
  \emph{ACM transactions on intelligent systems and technology (TIST)}, vol.~2,
  no.~3, pp. 1--27, 2011.

\bibitem{jing2019automatic}
S.~Jing, X.~Mao, and L.~Chen, ``Automatic speech discrete labels to dimensional
  emotional values conversion method,'' \emph{IET Biometrics}, vol.~8, no.~2,
  pp. 168--176, 2019.

\bibitem{toisoul2021estimation}
A.~Toisoul, J.~Kossaifi, A.~Bulat, G.~Tzimiropoulos, and M.~Pantic,
  ``Estimation of continuous valence and arousal levels from faces in
  naturalistic conditions,'' \emph{Nature Machine Intelligence}, vol.~3, no.~1,
  pp. 42--50, 2021.

\bibitem{yang2022hybrid}
L.~Yang, Y.~Shen, Y.~Mao, and L.~Cai, ``Hybrid curriculum learning for emotion
  recognition in conversation,'' in \emph{Proceedings of the AAAI Conference on
  Artificial Intelligence}, vol.~36, no.~10, 2022, pp. 11\,595--11\,603.

\end{thebibliography}


%
%
%
%


\end{document}